\documentclass[final,3p,times,twocolumn,authoryear]{elsarticle}

\journal{Icarus}

\begin{document}

\begin{frontmatter}

\title{KCTF Evolution of Trans-Neptunian Binaries:\\Connecting Formation to Observation}

\author[Lowell,SESE]{Simon B. Porter\corref{cor1}}
\ead{porter@lowell.edu}
\author[Lowell]{William M. Grundy}

\address[Lowell]{Lowell Observatory, 1400 W. Mars Hill Rd., Flagstaff, AZ 86001, USA} 
\address[SESE]{School of Earth and Space Exploration, Arizona State University, Tempe, AZ 85287, USA}

\cortext[cor1]{Principal corresponding author; Phone: 931-581-5441; Fax: 928-774-6296}

\begin{abstract}
Recent observational surveys of trans-neptunian binary (TNB) systems have dramatically increased the number of known mutual orbits.
Our Kozai Cycle Tidal Friction (KCTF) simulations of synthetic trans-neptunian binaries show that tidal dissipation in these systems can completely reshape their original orbits.
Specifically, solar torques should have dramatically accelerated the semimajor axis decay and circularization timescales of primordial (or recently excited) TNBs.
As a result, our initially random distribution of TNBs in our simulations evolved to have a large population of tight circular orbits.
This tight circular population appears for a range of TNO physical properties, though a strong gravitational quadrupole can prevent some from fully circularizing.
We introduce a stability parameter to predict the effectiveness of KCTF on a TNB orbit, and show that a number of known TNBs must have a large gravitational quadrupole to be stable.
\end{abstract}


\end{frontmatter}

\def\aap{Astron. Astrophys.~}
\def\apjl{Astrophys. J. Let.~}
\def\aj{Astron. J.~}
\def\icarus{Icarus~}
\def\nat{Nature~}
\def\apss{Astrophys. Space Sci.~}
\def\apj{Astrophys. J.~}
\def\mnras{Mon. Not. R. Astron. Soc.~}
\def\pasj{Publ. Astron. Soc. Jpn.~}


\section{Motivation}

Trans-neptunian binary systems (TNBs) constitute at least 10\% of the objects between 30 and 70 AU \citep{Stephens2006}, and up to 30\% of the Cold Classical Kuiper Belt \citep{Noll2008b}.
As of spring 2012, 72 TNBs have been reported in the literature, with full mutual orbits having been reported for 18 objects, 
partial orbits with ambiguous orbits for 30 more \citep[e.g.][and the list at http://www2.lowell.edu/users/grundy/tnbs]{Noll2008,Grundy2009,Grundy2011,Parker2011}.
These observations show that the majority of detected TNB systems have a separation of less than 2\% of the Hill Radius ($r_{Hill}$),
defined as:
\begin{equation}
\label{rHill}
r_{Hill} = a_{helio}(1-e_{helio})\sqrt[3]{\frac{M_{binary}}{3 M_{Sun}}}
\end{equation}
where $a_{helio}$ and $e_{helio}$ are the semimajor axis and eccentricity of the heliocentric orbit.
Even more striking is the very small fraction of TNB systems which are widely-separated ($>$10\% $a/r_{Hill}$), despite their being easier to detect.
This implies that TNBs are generally in very close mutual orbits, and the fraction of orbits that are very close has only increased with better detection methods.
In addition, most known TNBs are of almost equal brightness \citep{Noll2008}, implying near-equal masses.
\\\\
Several formation methods have been proposed to create TNBs, though none as yet can fully describe the observed population, nor account for any post-formation orbital evolution.
Large impacts are an obvious contender for formation, but tend to produce smaller satellites (and thus less equal mass ratios) than are observed.
Indeed, \citet{Canup2005} showed the Charon-forming impact required a very slow relative velocity ($v_{imp}\approx{v_{esc}}\approx0.7$ km/s), and even then only allowed a mass ratio of approximately 10:1.
Dynamical captures can also produce TNBs \citep[e.g.][]{Goldreich2002,Lee2007}.
These methods do favor near-equal mass ratios (as it provides a deeper gravity well per size of the primary object), but have great preference for producing wide ($>$5\% $r_{Hill}$) binaries on eccentric orbits.
\citet{Funato2004} combines a small impact and dynamical capture to efficiently produce TNBs, but only at very high eccentricities.
\citet{Nesvorny2010} shows that binaries formed by gravitational collapse also tend to have near-equal mass ratios, but again have wide, moderately eccentric orbits.
The unbinding of binaries by impacts \citep{Petit2004} or Neptune encounters \citep{Parker2010} would reduce the number of wide TNBs, but would not correspondingly increase the number of tight systems.
The deficit of these wide systems, and the abundance of tight ones, therefore hints at the existence of some non-disruptive post-formation processing of TNB mutual orbits.
\\\\
In this paper, we propose that Kozai Cycle Tidal Friction \citep[KCTF, after][]{Eggleton2006} may be the method by which these orbits were tightened and circularized.
We will show through several sets of Monte Carlo simulations that KCTF can transform a large fraction of primordial TNB systems into very close and circular orbits.
In addition, we show that the tidal evolution this implies means that Kozai cycles are very inefficient at destroying TNB systems.
We also show how KCTF is influenced by the physical properties of the TNB system, such as tidal $Q$ and $k_L$, density, $J_2$, rotation rate, and mass ratio.
\\\\
\textbf{Figure 1 here}

\section{KCTF Model}

\begin{figure*}[p!]
\centering
\includegraphics[width=10cm]{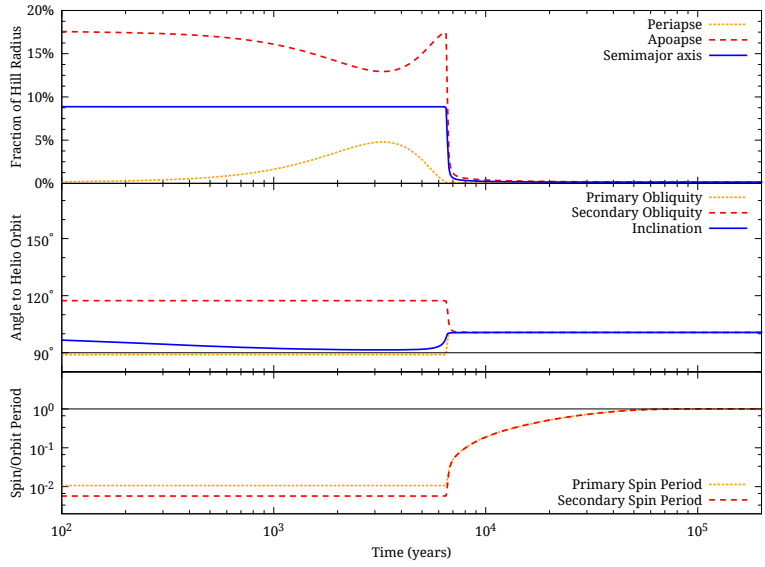}
\caption{An example of rapid KCTF evolution of a TNB mutual orbit; see Section 2.1 for a full description.}       
\label{fig:tale}
\end{figure*}

In order to understand how TNB orbits may have evolved since they were formed, we created a numerical Kozai Cycle and Tidal Friction model. 
Kozai Cycles in this context are periodic oscillations in eccentricity and inclination of the TNB mutual orbit caused by solar torques. 
For this paper, the outer orbit is the heliocentric orbit of the binary's barycenter, and the inner orbit is the mutual orbit of the binary pair.
These oscillations preserve the orbit's semimajor axis and the quantity $cos~I\times\sqrt{1-e_{in}^2}$, 
where $I$ is the inclination of the mutual orbit with respect to the heliocentric orbit and $e_{in}$ is the eccentricity of the inner orbit.
This process was first described in \citet{Kozai1962}, in the context of perturbations by Jupiter on asteroid orbits. 
Without any tidal or quadrupole effects, these oscillations would vary eccentricity periodically with a period between approximately 2 ka and 2 Ma. 
\citet{Kozai1962} showed that for an initially circular orbit, the minimum inclination for oscillation to initiate is $\pm$39.2$^{\circ}$. 
However, this limiting inclination becomes much lower at non-zero initial eccentricities. 
Thus, the only mutual orbits that could be excluded from this effect are those which form with both low initial eccentricity and low inclination relative to their heliocentric orbit.
Kozai cycles have been suggested as a method of evolving the mutual orbits of TNBs \citep{Perets2009}, as well as binary and triple Near-Earth asteroids, which may have Kozai cycles short enough to be observable \citep{Fang2011}.
\\\\
A significant consequence of these Kozai oscillations is that the eccentricity of the mutual orbit can become very high, especially if the initial orbit has a low eccentricity but high inclination (or vice versa).
Since the tidal dissipation rate for these objects is chiefly a function of their mutual separation at periapse \citep[see Equations 5 to 7 in][]{Eggleton2001},
a minor increase in eccentricity can have a major effect on the amount of tidal dissipation.
This is important, as tidal models that assume near-zero eccentricity would produce much slower tidal evolution than is realistic for an orbit with high eccentricity due to solar Kozai effects.
Mutual orbits with Kozai-pumped eccentricities can therefore decay due to tidal friction much faster than their initial state would imply; see Figure \ref{fig:tale} for an example. 
The strength of Kozai-driven tidal decay is inversely proportional to the magnitude of the binary orbit's angular momentum as projected on the axis of the heliocentric orbit's angular momentum vector.
Also, because this projected angular momentum is perpendicular to the solar-driven precession of the system, it can be completely determined from the instantaneous orbit without knowledge of the precession.
Here, we normalize this quantity to a percent of the Hill radius of the system:
\begin{equation}
\label{Hp}
H' = cos~I\sqrt{a_{in}(1-e_{in}^2)\frac{100}{r_{Hill}}}
\end{equation}
Where $a_{in}$ is the semimajor axis of the inner orbit.
We find this to be a particularly useful normalization, as values of $H'$ smaller than one will experience strong body tides over the course of a Kozai cycle, 
while larger values generally will not (depending on their physical properties).
Also, since we do not evolve the binary's heliocentric orbit in our simulations, 100/$r_{Hill}$ is a constant normalization parameter.
$H'$ is effectively Tisserand's Parameter for a three-body system with only quadrupole perturbations, which is appropriate here because all known TNBs have 
$a_{helio}{\gg}a_{in}$ by at least five orders of magnitude.
\\\\
Since $H'$ is a much stronger function of the orbit's orientation than separation, even very wide binaries can be affected by KCTF if their inclination (or eccentricity) is high enough.
This KCTF process has been previously identified as significant for TNBs by \citet{Perets2009}, but only demonstrated for the Orcus-Vanth dwarf planet system \citep{Ragozzine2009}.
We used a similar model based on \citet[][herein EKE01]{Eggleton2001} to \citet{Ragozzine2009}, which is described below.
This model directly evolves the mutual orbital elements and spin vectors of the binary while holding the heliocentric orbit constant. 
We did not include any dynamical effects from objects external to the binary other than the Sun. 
The general equations of the EKE01 KCTF model are summarized in \citet{Fabrycky2007}; below we describe the modifications and additions we used.
These consist of our estimation of frictional timescale, quadrupole gravity terms, and integration methods.

\subsection{Frictional Timescale}

Since the EKE01 model was developed for binary stars and giant planets, we needed to modify the terms relating to the physical characteristics of the objects.
One benefit of the EKE01 method is that all these terms are condensed into a single frictional timescale for each object.
This timescale, however, is also a function of the mutual orbit's semimajor axis and is thus time-varying.
We therefore reformulated the frictional timescale in a way that is computationally more useful.
\\\\
In the EKE01 model, the behavior of the objects' body tides is determined by the second tidal Love number ($k_L=k_2$) and the tidal dissipative function ($Q$) for each object.
The Love number is highly dependent on both the composition of the object and whether it is physically a solid object or a rubble pile.
For half of our simulations, we assumed the objects were solid homogeneous elastic bodies \citep{Burns1977}:
\begin{equation}
\label{kL1}
k_{L,solid}=\frac{3}{2} \left( 1+\frac{19 \mu_{r} R}{2 G M \rho} \right) ^{-1} 
\end{equation}
We took the rigidity of the objects, $\mu_{r}$, to be $4 \times 10^9 N/m^2$, using the value for icy bodies from \citet{Gladman1996}.
For the other half of the simulations, we assumed the objects were rubble piles, using the approximation of \citet{Goldreich2009}:
\begin{equation}
\label{kL2}
k_{L,rubble}=\frac{R}{10^5~\textrm{km}}
\end{equation}
In addition, we assumed a constant value for $Q$, the inverse of the average fraction of tidal energy lost to heat per radian of the orbit \citep{Goldreich1966}.
We can then find the tidal timescales as a function of the binary's orbit ($a_{in},n_{in}$), $k_L$, and $Q$. 
The viscosity ($t_V$) and frictional ($t_F$) timescales for the primary object as were formulated in \citet[][Equations A9 and A10]{Fabrycky2007} as:
\begin{eqnarray}
t_{V,1} &=& \frac{3}{2} \frac{(1+k_{L,1})^2}{k_{L,1}} \frac{Q_1 n_{in} R_1^3}{GM_1} \\
t_{F,1} &=& \frac{t_{V,1}}{9} \left(\frac{a_{in}}{R_1}\right)^8 \frac{M_1^2}{(M_1+M_2)M_2} (1+k_{L,1})^{-2}
\end{eqnarray}
The timescales for the secondary object are the same equations, but switch the subscripts 1 and 2.
Combining these two equations allows $a_{in}$ to be separated out, thus reducing the number of calculations required per iteration:
\begin{eqnarray}
\label{eq:tF}
t_{F,1} &=& \frac{1}{6}\frac{Q_1}{k_{L,1}}\frac{M_1}{M_2}\frac{R_1^{-5}}{\sqrt{G(M_1+M_2)}} \times a_{in}^{13/2}
\end{eqnarray}
Note the leading factor of $1/6$ is erroneously listed as $2/3$ in \citet{Ragozzine2009}, leading to longer frictional timescales, and thus slower orbital decay.
In addition, this timescale is for an object in a perfectly circular orbit with its rotation synchronized to the orbit.
Equations 5 and 6 in EKE01 combine these factors to account for the eccentricity of the orbit and the rotation of the objects.
Since the closest separation of the two objects is the key driver for tidal evolution, the effective timescale is very sensitive to eccentricity.
\\\\
This frictional timescale gives the approximate rates of evolution for a near-circular orbit, but the actual rates are strongly dependant on the orbit's eccentricity.
To illustrate this point, consider the special case of a system where the objects have equal values of $t_F$, have $e\gg0$, and their rotation is synchronized to the orbit, 
the rates of change for the semimajor axis and eccentricity can be approximated as:
\begin{eqnarray}
\label{eq:rates}
\frac{da}{dt} &=& \frac{-a}{t_F (1-e)^{15/2}} \\
\frac{de}{dt} &=& \frac{-1}{t_F (1-e)^{13/2}}
\end{eqnarray}
While the eccentricity is $e\gg0$, the semimajor axis decay can then be estimated as:
\begin{eqnarray}
\frac{a}{a_0} &=& \frac{t_{F,0}(1-e_0)^{15/2}}{t+t_{F,0}(1-e_0)^{15/2}}
\end{eqnarray}
Where $a_0$, $e_0$, and $t_{F,0}$ are the initial semimajor axis, eccentricity, and frictional timescale.
As an example, consider an equal-mass TNB system with objects having a radius of 100 km, $Q$=100, rubble-pile $k_L$, and density of 1.0 $g/cm^3$.
At a semimajor axis of 10$^4$ km (100 radii), the objects would have values of $t_F\approx7\times10^{12}$ years.
If the initial eccentricity were 0.5, the orbit would only have decayed to 9998 km after one million years.
At an eccentricity of 0.8, the semimajor axis would decay to 8512 km in that time.
And at an eccentricity of 0.9, the orbit would shrink to just 306.5 km (3 radii) after a million years.
By the time the system reaches an orbit this tight, $t_F$ has decreased to less than 1000 years, allowing for rapid circularization.
Clearly then, only a brief excursion to high eccentricity is needed to start a feedback loop of tidal decay to a tight circular orbit.
In KCTF, those excursions happen when Kozai cycles pump up the eccentricity.
\\\\
The system in Figure \ref{fig:tale} shows this process in action.
The two objects are equal mass rubble-piles with Q=10 and radii of 42 km.
The initial orbit is at 9.7\% of $r_{Hill}$, $e$=0.99, and inclination of 99$^\circ$.
However, the system's orientation puts it initially past the peak of its Kozai cycle, and so the orbit starts to become more circular and less inclined with little tidal evolution.
After reaching a minimum of $e$=0.46 at 3200 years, the eccentricity then grows again.
By the time it reaches $e$=0.98, the semimajor axis is shrinking at a rate of 2 km/year and 45 km/year at $e$=0.983.
The shrunken semimajor axis reduces $t_F$, so the peak decay rate is 150 km/year at 6550 years since the start, $e$=0.985, and $a$=6.1\% of $r_{Hill}$.
This corresponds to a periapse distance of 212 km, or 5 radii apart.
The orbit then gradually decays down to a tight circular orbit at 0.15\% $r_{Hill}$ or 8.1 radii.
\\\\
Most KCTF systems do not evolve as rapidly as this, and will often have several eccentricity peaks with a small amount of decay until the periapse becomes close enough for rapid decay to occur.
Still, a large range of orientations can produce strong KCTF evolution.
To measure this effect, we reran a number of the simulations described below without any solar perturbations.
In all these cases, the number of systems which circularized was reduced by at least a factor of 3.5 compared to the full KCTF model.
These systems all oriented randomly on the sphere of the sky (corresponding to observed systems), 
so the joint process of KCTF can cause fast tidal evolution in a large variety of TNB mutual orbits.
\\\\
\textbf{Figure 2 here}

\subsection{Quadrupole Gravity}

\citet{Ragozzine2009} expanded this model by adding the capacity for the objects to have a permanent quadrupole term in their gravity field. 
The non-uniform gravity field this allows is more physically appropriate for solid objects (like TNOs) than the stars and giant planets for which the EKE01 model was developed. 
This non-uniformity is especially relevant for objects the size of most known TNBs (less than 400 km diameter), 
which are likely not large enough to have reached hydrostatic equilibrium \citep{Yasui2010}, and thus could have relatively large quadrupole fields. 
The quadrupole moment is defined for an axisymmetric body as $J_2$=$(C-A)/(M R^2)$, where $C$ is the moment of inertia about the polar radius, 
$A$ is the moment of inertia about the equatorial radius, $R$ is the equatorial radius, and $M$ is the mass.
In the EKE01 model, the vector $(X,Y,Z)$ provides the angular precession rate relative to the inertial frame, 
and is in the $(\hat{e}_{in},\hat{q}_{in},\hat{h}_{in})$ orthonormal basis, 
where $\hat{e}_{in}$ is the normalization of the mutual orbit's Laplace-Runge-Lenz vector, which points in the direction of the periapse, 
$\hat{h}_{in}$ is in the direction of the orbit's angular momentum vector, and $\hat{q}_{in} = \hat{h}_{in} \times \hat{e}_{in}$.
This vector due to solar torques is given by Equations 10 to 12 in EKE01, and their relation to the secular evolution of the orbit is summarized by Equations A6 to A8 in \citet{Fabrycky2007}.
\citet{Ragozzine2009} formulated the additional precession due to the primary's quadrupole field as being:
\begin{eqnarray}
X_{J_{2,1}}&=&\frac{3}{2}J_{2,1}\left(\frac{R_1}{a_{in}}\right)^2\frac{n_{in}}{(1-e_{in}^2)^2}\frac{\Omega_{1h}\Omega_{1e}}{\Omega_1^2} \\
Y_{J_{2,1}}&=&\frac{3}{2}J_{2,1}\left(\frac{R_1}{a_{in}}\right)^2\frac{n_{in}}{(1-e_{in}^2)^2}\frac{\Omega_{1h}\Omega_{1q}}{\Omega_1^2} \\
Z_{J_{2,1}}&=&\frac{3}{4}J_{2,1}\left(\frac{R_1}{a_{in}}\right)^2\frac{n_{in}}{(1-e_{in}^2)^2}\frac{2\Omega_{1h}^2-\Omega_{1e}^2-\Omega_{1q}^2}{\Omega_1^2}
\end{eqnarray}
Where the terms $a_{in}$, $e_{in}$, and $n_{in}$ are the semimajor axis, eccentricity, and mean motion of the mutual orbit, and $R_1$ is the radius of the primary. 
In addition, $\Omega_{1i}$ is the projection of primary's spin angular velocity vector onto the axis $i$. 
Since the binaries dealt with in this work are of similar size, this process was repeated to account for the secondary's quadrupole field, 
and the two quadrupole precession vectors added on to the $(X,Y,Z)$ vectors defined in the EKE01 model.
The total $(X,Y,Z)$ vector was then combined with the dissipative terms (our Equation \ref{eq:tF}) to feed the full evolution equations (Equations 1 to 6 in EKE01).

\subsection{Integration Methods}

Since the EKE01 model defines the evolution of the system by a set of four related inhomogeneous vector differential equations, 
we needed a numerical integrator that could solve them rapidly and precisely on a modern computer.
For this we based our integrator on a Burlisch-Stoer method, which combines a modified-midpoint integrator with an polynomial interpolation method to increase precision and control error.
Since the system was conservative when the dissipative terms where close to zero, 
we used a fixed timestep of 1.1 mutual orbital periods when $|V_1|+|V_2|<10^{-18}sec^{-1}$ (equating to an approximate circularization timescale of $3.2{\times}10^{10}$ years).
On the other hand, if the system were dissipative, we used the adaptive timestep management algorithm described in \citet{Press2007}, setting a minimum timestep of 10 days (864000 sec).
This algorithm estimated the total error for each step as the root-mean-square of the normalized error estimates for each component 
($e_{in},~\hat{e_{in}},~h_{in},~\hat{h_{in}},~\vec{\Omega_1},~\vec{\Omega_2}$), and kept it below a tolerance of $10^{-13}$ for each timestep.
Throughout the simulation, the program keeps track of the total angular momentum (spin plus orbit) in the direction of the heliocentric orbit's angular momentum vector.
As described above, this term determines the magnitude of solar perturbations on the mutual orbit and should be precisely preserved over the entire integration.
The small number of cases (generally those with non-zero values of $J_2$) which did not preserve angular momentum were rerun at sufficiently smaller tolerances that momentum was again conserved.
Conversely, if this tight tolerance proved to be numerically unstable, the tolerance was slowly increased until it was stable but still conservative.
\\\\
We ended each simulation when it reached either 4.5 Ga (i.e. the maximum physically possible evolution time) or reached an eccentricity smaller than 10$^{-4}$.
This value of minimum eccentricity was chosen for an end state because preliminary simulations down to 10$^{-10}$ showed no further orbital evolution beyond circularization.
In addition, the simulation would end prematurely if the periapse fell below the Roche limit ($\approx$1.26($R_1+R_2$), which we consider an impact),
the apoapse grew beyond the Hill radius of the system, or one of the objects reached a spin period faster than the breakup rotation rate.
The condition for the latter case was a rotation rate greater than $2\pi\sqrt{G\rho/(3\pi)}$, and in practice was never reached in our simulations.

\section{Monte Carlo Simulations}

To test the responses of TNBs to the KCTF model, we conducted a series of Monte Carlo simulations in which we created a sample set of 1000 synthetic TNBs with randomized mutual orbital elements and system masses.
The heliocentric orbit, physical properties, and range of rotation rates were all kept constant for each set. 
We then evolved each system for 4.5 billion years, or until the system either circularized, impacted itself, became unbound, or spun to breakup.
\\\\
As common initial parameters, we set the heliocentric orbits to a semimajor axis of 45 AU and eccentricity of 0.05, representative of the cold-classical belt, which contains the highest fraction of TNBs.
We then varied the system GM range from 0.02 to 0.20 km$^3$/s$^2$. 
This corresponds to a radius range from 33 to 71 km for an equal-mass, $\rho=1.0~\textrm{g/cm}^3$ system, appropriate for the lower range of detectable TNB systems.
The semimajor axis of the mutual orbit (in the frame of the primary) was varied uniformly from 0.1\% to 10\% of the system's Hill radius.
This range is inclusive of nearly all known TNB orbits, as well as published formation models.
The mutual eccentricity was varied uniformly from 10$^{-4}$ to 0.9999.
The orbits were all orientated randomly on the sky by first generating a random direction for the $\hat{h_{in}}$ vector with the algorithm of \citet{Knop1970}.
A second random vector perpendicular to the first was then generated and used as the $\hat{e_{in}}$ vector.
These two vectors plus the semimajor axis, eccentricity, and system mass then fully describe the randomized orbit.
This random orientation corresponds to dynamical simulations of binary capture \citep{Kominami2011}.
Likewise, the initial spin poles of the two objects were pointed at two different vectors also randomized on the sky.
\\\\
We considered two different densities, 0.5 and 1.0 g/cm$^3$, representative of the range of reported densities for binary TNOs in this size range \citep{Stansberry2011}.
As noted above, we also considered both solid and rubble-pile assumptions for $k_L$.
For each simulation set, we replicated the runs for each of the four combinations of density and $k_L$.
For most of the simulations, we allowed the initial rotational periods of the objects to vary between 4-48 hours.
This range is consistent with observed lightcurves for solitary TNOs \citep{Trilling2006,Thirouin2010}.
To check the sensitivity of this range, we also ran simulations with 2-7 day rotation rates.
We ran most of our simulations with $J_2$=0, but we also ran sets at $J_2$=0.01 and $J_2$=0.1.
The former being comparable to Saturn's satellite Phoebe, and the later to Hyperion, both of which are in the same size range as our simulations.
We ran most simulations with $Q$=100, a typically canonical value for large solid objects.
However, since smaller objects may have much smaller $Q$ \citep{Goldreich2009,Zhang2008}, we also performed a set of simulations with $Q$=10.
Finally, in addition to the equal-mass simulations, we ran a set with a mass ratio of 10:1.
Since most observed binaries have a brightness difference of less than one magnitude \citep{Noll2008}, these two mass ratios are representative of the observed population.
\\\\
The parameters varied per each set of 1000 simulations were thus $Q$, $J_2$, rotation rate, mass ratio, density, and $k_L$. 
Table \ref{tab:sims} lists the 24 sets considered, for a total of 24,000 simulations and approximately 1 million CPU-hours.
Because the initial conditions of these systems were distributed across reasonable ranges of $a$, $e$, $i$, and obliquity, they do not necessarily represent the primordial population of TNBs.
Rather, they are a superposition of all the possible initial states.
KCTF can then act as a filter to check whether a certain range of initial orbits (and thus formation methods) corresponds to the observed population of TNBs.
\\\\
\textbf{Figure 3 here}
\\\\
\textbf{Table 1 here}

\section{Results}

\begin{figure*}[p!]
\centering
\includegraphics[width=10cm]{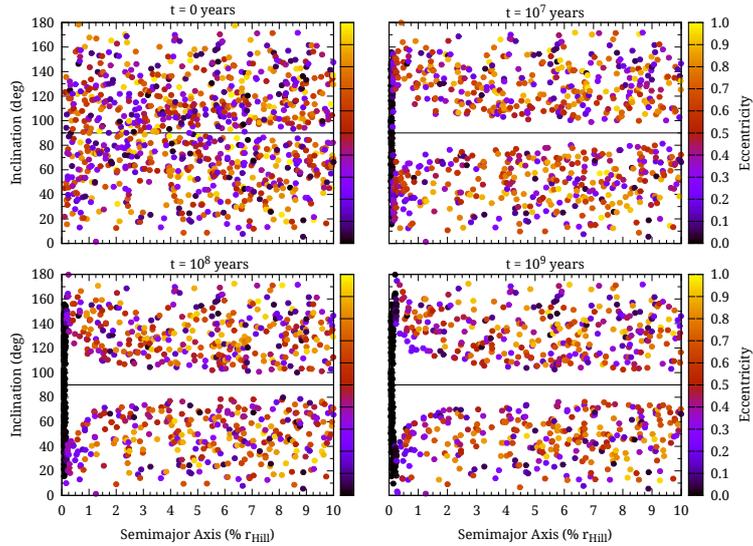}
\caption{The evolution over time of 1000 synthetic equal-mass TNBs with $Q=100$, $\rho=1.0$ g/cm$^3$, elastic solid $k_L$, and $J_2=0$.}
\label{fig:seq}
\end{figure*}

\begin{figure*}[p!]
\centering
\includegraphics[width=10cm]{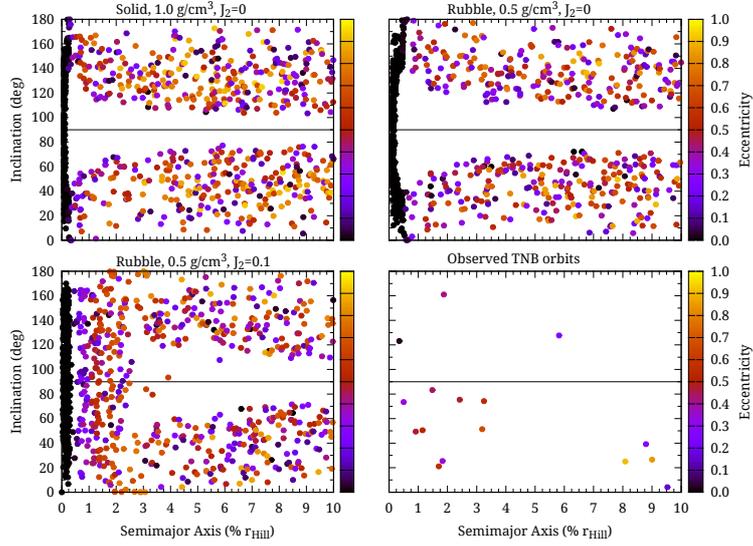}
\caption{A comparison of end states for three different physical properties and observed orbits. The simulations are all equal-mass, $Q=100$. 
Observed orbits are as listed in Table \ref{tab:obs}.
}       
\label{fig:icomp}
\end{figure*}

\begin{figure*}[p!]
\centering
\includegraphics[width=10cm]{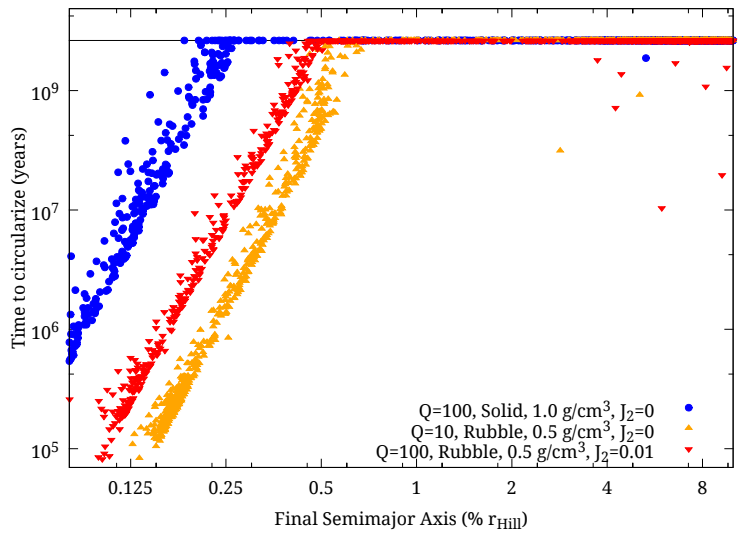}
\caption{The circularization times for three different physical properties; the simulations are all equal-mass systems. 
The horizontal line corresponds to 4.5 billion years, and any points along it represent stable elliptical orbits.
The points on the left below the line are simulations that were evolved by KCTF to close circular orbits,
while the points on the right below the line started at very low eccentricity and evolved to $e<10^{-4}$.
The final semimajor axis in tidally-evolved systems is usually very close to the periapse of the orbit when body tides become the dominant force,
as most of the energy loss in an evolving orbit is at the periapse.}
\label{fig:circ}
\end{figure*}

We found that a significant fraction of our simulations resulted in the synthetic TNB systems evolving to very tight circular orbits; Table \ref{tab:sims} lists the relative fractions for each simulation.
These tight, circular orbits were at less than 1\% of $r_{Hill}$ and eccentricities smaller than 10$^{-5}$, meaning that they were entirely dominated by mutual tidal interactions.
In addition to these highly-evolved systems, a number of systems evolved to orbits that were tighter but still eccentric.
Figure \ref{fig:seq} shows the time evolution of one set of simulations and Figure \ref{fig:icomp} separates initial and final states for the circularized and elliptical cases of the same set.
The conservation of $H'$ is quite evident in these plots, as is its dependence on $cos~I$ and $\sqrt{a}$.
Figures \ref{fig:circ}, \ref{fig:Hp}, and \ref{fig:ob} show the final states of a range of different simulation sets, varying in both the speed of body tides and physical shape.
\\\\
Much of the tightly-bound population is beyond current TNB detection methods, including the WFC3 camera on the Hubble Space Telescope and the laser guide star adaptive optics system at the Keck Observatory.
However, some of the known TNO orbits do fall within this population.
The closest published full orbit is (79360) Sila-Numan, formerly 1997 CS$_{29}$, at 0.35\% $r_{Hill}$ \citep{Grundy2012}.
The published eccentricity is 0.02, but a fully circular orbit is only excluded at 1.8 $\sigma$ confidence.
In addition, (120347) Salacia-Actaea at 0.23\% $r_{Hill}$ \citep{Stansberry2011} and the centaur (65489) Ceto-Phorcys at 0.46\% $r_{Hill}$ \citep{Grundy2007} both have extremely circular orbits.
These three objects likely represent the inner edge of the potentially very numerous tight circular population, and should become less unique as observations improve.
\\\\
Our simulations also naturally produced a deficit of systems with a final semimajor axis smaller than $1.26\times(R_1+R_2)$, our approximate Roche limit.
It is perfectly possible that these systems could survive, either as a contact binary or breaking apart and reforming in a more stable configuration \citep{Jacobson2011}.
However, any of these cases are beyond the capabilities of our model and would require further work.
This inner edge corresponds to a $J/J'$ of approximately 0.4, where $J$ is the total orbit + spin angular momentum, $J' = \sqrt{GM_t^3R_{eff}}$, $M_t$ is the system mass, 
and $R_{eff}$ is the radius of a sphere of mass $M_t$ and density equal to the average density of the two objects.
\citet{Canup2005} showed that binaries with $J/J' < 0.8$ can be produced by collisions.
Thus, we found that KCTF can transform a wide range of binaries formed by capture into ones sufficiently close as to be indistinguishable from collision-produced binaries, 
the general case of what was shown for Orcus-Vanth by \citet{Ragozzine2009}.
\\\\
Because the quadrupole component of Kozai perturbations are axisymmetric, the resulting systems also preserve the direction of their initial mutual orbit relative to the plane of their heliocentric orbit. 
The EKE01 model only includes the quadrupole component, though in the actual dynamics of these systems there are also higher-order terms.
The next higher term is the octupole component, which is not axisymmetric and can therefore cause a system to flip from prograde to retrograde \citep[and vice versa, ][]{Naoz2011}.
However, the relative strength of the octupole to the quadrupole goes as $a_{in}/a_{helio}$, and so the quadrupole completely dominates in all the cases we considered \citep{Ragozzine2009}.
Thus, the initial prograde/retrograde ratio was preserved in all our simulations. 
\\\\
\textbf{Figure 4 here}

\subsection{Stability of Orbits to KCTF}

\begin{figure*}[p!]
\centering
\includegraphics[width=10cm]{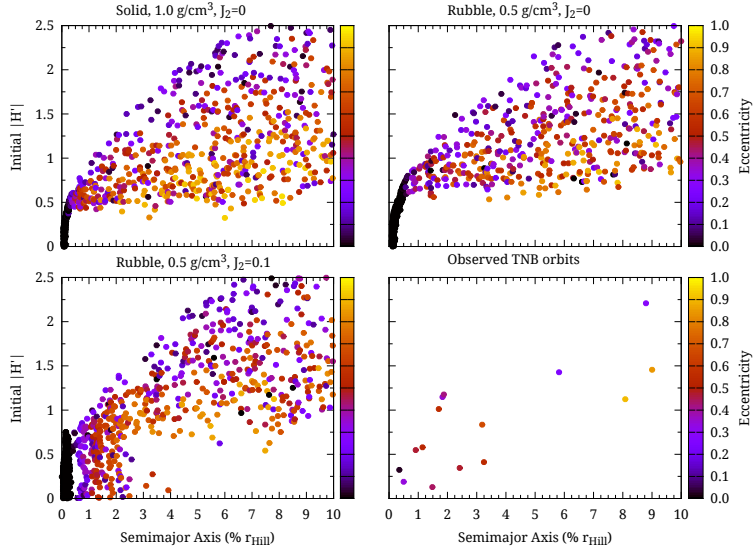}
\caption{A comparison of initial $|H'|$ and final semimajor axis for three different physical properties and observed orbits. The simulations are all equal-mass, $Q=100$. 
Observed orbits are as listed in Table \ref{tab:obs}.
}
\label{fig:Hp}
\end{figure*}

A further inclination effect can be seen in Figure \ref{fig:icomp}; the inclination region within 10$^{\circ}$ of perpendicular to the heliocentric orbit is empty for all but the tightest semimajor axes.
This range of inclination equates to such low values of $H'$ that all orbits starting there reach very high levels of eccentricity, and therefore initiate runaway KCTF decay (e.g. Figure \ref{fig:tale}).
Figure  \ref{fig:Hp} clearly shows that these orbits mostly end up tighter than 1\% of $r_{Hill}$.
Indeed, as Figure \ref{fig:Hp} shows, there is a clear limiting value of $|H'|$ below which orbits are not stable to KCTF.
\\\\
We define this limiting value, $|H'_{tide}|$, as the minimum value of $|H'|$ where the relative difference in the initial and final semimajor axes is less than 10\%, i.e. $|a_i-a_f|/a_i<0.1$.
The values of $|H'_{tide}|$ for all our simulations are listed in Table \ref{tab:sims}.
The average values of this $|H'_{tide}|$ are all close to 1.0, with the higher-dissipation cases above and the lower below.
Thus, $|H'|$ appears to be a good normalized indicator of KCTF susceptibility for TNB orbits; values below 1.0 likely will have experienced Kozai cycle-driven tidal evolution, while values above will likely have not.
This is especially useful for determining the stability of observed binary systems where the mutual orbit is known, but not any physical or rotational properties.
\\\\
In addition, we found it useful to define $|H'_{circ}|$, the minimum value of $|H'|$ for an orbit that did not circularize.
This value ranges from 0.2-0.7 for non-$J_2$ runs, but is close to zero for the simulations with $J_2$.
This shows that $J_2$ can provide an island of stability for eccentric orbits close enough that $J_2$ blocks further Kozai cycles, but not close enough for further tidal decay.
These systems have tidally evolved beforehand to reach this state, but are stable once they reach it.
\\\\
\textbf{Figure 5 here}

\subsection{Effects of Physical Parameters}

For our base sets of simulations, we assumed $Q$=100, $J_2$=0, rotation rates between 4 and 48 hours, and equal masses for the two objects.
Using these parameters, we ran sets at each combination of density equals 0.5 or 1.0 g/cm$^3$ and a Love number appropriate for either a rubble pile or elastic solid body.
The remainder of the simulations perturbed one of the first set of parameters and ran the same set of four density/Love number combinations; Table \ref{tab:sims} shows this grouping.
\\\\
In general, the low-density rubble piles were the most susceptible to tidal decay, followed by the high-density rubble piles.
This makes sense, as the internal friction of a rubble pile allows it to dissipate tidal forces very effectively.
The lower density allowed a larger radius per mass, thus raising larger tides and a higher Love number (by Equation \ref{kL2}).
Less obvious is that fewer of the low-density elastic solid simulations decayed to stable circular orbits than the high-density cases.
Here, the difference can be seen in the final column of Table \ref{tab:sims}; these low-density, rigid systems suffered a much higher rate of mutual collisions.
Their larger radius prevented a significant fraction of the very eccentric systems from circularizing, lowering the overall efficiently of producing close, circular orbits.
As noted below, these mutual collisions could still produce close or contact binaries, but that is beyond the capabilities of these simulations.
\\\\
While we assumed the canonical value of tidal $Q$=100 for most of the simulations, a lower value is likely more physical for the size of objects we considered \citep{Goldreich2009}.
The simulations that we ran with $Q$=10 did show an average of 20\% greater propensity to tidally decay and circularize.
This small increase in tidal decay for a full order of magnitude decrease in $Q$ shows the true driver in this evolution is the closest periapse the system reaches.
Either the objects become close enough to undergo decay or they do not, there is not much space in between.
Indeed, the increase is roughly comparable (100/10)$^{1/8}$, where the strength of the tides goes as $a^8$.
An additional effect of the lower $Q$ is to remove the higher mutual collision probability for the low-density, elastic solid case.
As shown in Figure \ref{fig:circ}, the faster tides allow these cases to circularize at larger separations, limiting the impact probability.
\\\\
We ran sets of simulations with a $J_2$ of both 0.01 and 0.1.
As seen in Figure \ref{fig:ob}, the main effect of $J_2$ on wider systems was to decrease the obliquity of the objects with respect to their mutual orbit.
For slightly more evolved systems, though, an interesting effect could be seen;
a number of the systems decayed to a point where $J_2$ was strong enough to block further Kozai cycles, but their periapse was wide enough at the point of being frozen that no further tidal decay could occur.
This freezing-in of Kozai-blocked systems created the island of high-inclination, moderately eccentric systems seen in Figures \ref{fig:icomp} and \ref{fig:Hp}.
Since the frozen systems did not evolve further, they reduced the total fraction of systems that circularized.
However, since they had to first tidally evolve to reach the frozen state, the total fraction that tidally evolved (and thus $|H'_{circ}|$) is almost the same as the non-$J_2$.
Also, since tidal evolution is strongly driven by the periapse, there is a noticeable gradient in eccentricity, with the closest frozen systems having $e<0.2$. 
Thus, the effect of $J_2$ on KCTF of TNBs is produce stable high-inclination, moderately eccentric TNBs, while not eliminating the tight circular population.
\\\\
We also ran a group of simulations with much slower initial rotation rates, from 2-7 days (and at random initial orientations, just as before).
These simulations were slightly less likely to evolve to close circular orbits than their fast rotating counterparts.
This may be because the final orbital periods of the close circular orbits were closer to the initial rotation rates of the faster rotators.
Thus, the slower rotators required, on average, a larger amount of spin-orbit interaction to reach a doubly-synchronous state (with both spin periods equal to the orbital period).
The effect was slight enough, though, to conclude that initial rotation rate is not a significant factor in the evolution of these TNBs.
\\\\
Finally, we changed the mass ratio of the two objects to be 10:1.
Since most known TNBs have near-equal brightness ratios \citep{Noll2008}, this covers most of known TNBs that are not dwarf planetary systems.
The effect on tidal decay was again similar to the slower initial rotation rates, a slight decrease in the efficiency of producing close circular orbits.
Here the reason is that most of the dissipation is taking place on the secondary, which can only have smaller tides due to its smaller radius.
Again, though, the effect is slight enough that most systems would be completely insensitive to mass ratios between 1:1 and 10:1.
\\\\
\textbf{Table 2 here}

\subsection{Survival Probability}

Not all of the simulated binaries survived to either circularize or reach 4.5 Ga.
As shown in Table \ref{tab:sims}, about 1-15\% of our random initial orbits proved in some way unstable.
The largest fraction of destroyed systems was due to impacts when Kozai Cycles drove the periapse of the system within the mutual Roche radii of the objects.
This was especially apparent for systems with elastic solid values of $k_L$ (weak tides) and large radii, which consistently had destruction rates higher than 10\%.
Every other case had destruction rates smaller than 10\%.
As noted above, this is because the low-density solid systems circularize slowly enough that semimajor axis decay can bring their periapse below the Roche limit.
In real TNBs, such a close approach could result in the temporary breakup of the secondary and reformation of one or more new satellites in tight circular orbits.
The simulated systems were also considered destroyed if the apoapse of the system exceeded $r_{Hill}$, though this case did not occur in any of our simulations. 
Likewise, we had no systems spin to breakup, with spins either slowing to synchronize or not changing at all (see Figure \ref{fig:ob}).
It therefore appears that mutual collisions are the only practical way to destroy a TNB system with KCTF, and even then only with an efficiency of a few percent.
In addition, 90\% is a lower limit for the number of TNBs (of an initially random population) which survive as binaries once formed, exclusive of impacts or close encounters.
\\\\
The survival rate in our simulations is much higher than that calculated by \citet{Petit2004}, who found that non-disruptive impacts were the most effective means of destabilizing wide TNBs.
This implies that small impacts and disruption by Neptune encounter \citep{Parker2010} are still the most efficient means to cause a TNB to become unbound.
However, our results show that KCTF is very efficient at transforming wide binaries into tight ones, which would presently be observed as single objects.
Indeed, KCTF can quickly shrink wide binaries down to orbits tight enough to have a high survival probability in the event of a Neptune encounter \citep{Parker2010}.
Thus, any scaling of the initial population of wide TNBs based on survival rates must also account for the fraction which decay into much tighter orbits.
\\\\
\textbf{Figure 6 here}

\subsection{Obliquity}

\begin{figure*}[p!]
\centering
\includegraphics[width=5cm]{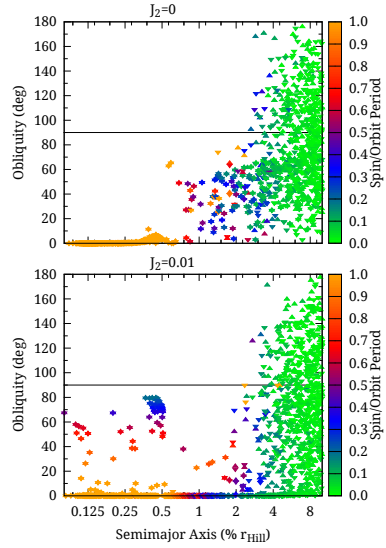}
\caption{The final spin rate and obliquity (to the mutual orbit) with and without $J_2$; upward-triangles are the primary objects, and downward are the secondaries. 
The high obliquity of some of the close orbits with $J_2$ is due to them being captured into Cassini state 2 (see text).} 
\label{fig:ob}
\end{figure*}

All the systems we simulated started out with randomly directed spin axes, and thus each object had random initial obliquities.
We define obliquity here to be the angle between the spin pole of the objects and a vector perpendicular to the plane of the mutual orbit.
Thus, an obliquity of 0$^{\circ}$ is parallel to the orbit's angular momentum vector, while an obliquity of 90$^{\circ}$ is perpendicular.
Figure \ref{fig:ob} shows the final obliquities of equal-mass, $Q$=100 simulations with and without $J_2$.
\\\\
In general, the non-$J_2$ runs experienced two stages of obliquity evolution.
At less than about 3\% $r_{Hill}$, the spins of the two objects begin interacting with each other and with the orbit.
This causes any objects rotating retrograde to their orbit to flip around to prograde, causing the step jump in Figure \ref{fig:ob}.
In addition, the two objects begin to match their obliquities and rotation rates to each other, thus producing the star shapes in Figure \ref{fig:ob}.
The second (and terminal) stage is for the objects to match their spin poles and rotation rates to their mutual orbit.
This drive to zero obliquity only occurs for the non-$J_2$ simulations when tidal forces are also strong enough to circularize the orbit.
\\\\
The simulations with $J_2$ were similar, but the additional gravitational factor from $J_2$ caused them to go to zero obliquity much faster.
Indeed, even some very wide orbits with little tidal evolution were still driven to zero obliquity by $J_2$.
Shape effects thus can have a major effect on the rotational evolution of even wide systems.
\\\\
A further effect can be seen in Figure \ref{fig:ob}, as some circularized orbits ended with non-zero obliquities and non-synchronous rotation rates.
This effect was more pronounced for less-dissipative simulations.
These systems appear to have been captured into Cassini state 2, which is low obliquity for fast precession and high obliquity for slow precession \citep{Fabrycky2007b}.
As they circularize at near-constant semimajor axes, their precession rate due to $J_2$ drops (equations 11-13).
Since they are sufficiently close that $J_2$ precession dominates, the overall precession rate becomes very small and the objects follow Cassini state 2 to high obliquity.
\\\\
This capture process is not necessarily observable in real systems.
Most small objects with known shapes (asteroids and minor satellites) are not pure oblate spheroids, but rather more complex shapes.
It is possible for these shapes to capture into Cassini state 2 \citep{Peale1969,Bills2008}, but it is much more difficult.
In addition, they may have precession rates due to higher-order terms that are large enough to prevent even a very circular system in Cassini state 2 from going to high obliquity.
Thus, a much more through study of the possible post-circularization rotational evolution of TNBs is needed to properly predict their current states.
\\\\
\textbf{Table 3 here}

\section{Discussion}

KCTF provides an evolutionary path to convert wide, elliptical binaries into close, circular ones.
It therefore helps to explain the observed dichotomy between the few (but well-sampled) wide TNB systems and the apparently numerous tight systems.
In addition, because it is a process that does not require any external forces other than the Sun, it applies to all objects that could be called trans-neptunian binaries,
from classical Kuiper Belt objects to highly scattered centaurs like Ceto-Phorcys.
And, because it is independent of the surrounding disk, KCTF can shrink and circularize orbits over a much longer timescale than disk dynamical friction (e.g. $L^2S$).
\\\\
A consequence of this extensive evolution is that KCTF should have significantly reshaped the orbits of most TNBs since their formation.
In the process, some information is lost, as tidal decay is an irreversible thermodynamic reaction.
The distribution of semimajor axes for present-day TNBs is thus necessarily tighter than at formation.
By what factor it is tighter is hard to determine, as the physical properties of the objects do affect the efficiency of semimajor axis decay (see Table \ref{tab:sims}).
Similarly, the exact eccentricity and inclination of the orbit are changed in ways that have partially erased their history.
\\\\
More clear is that KCTF generally preserves the prograde/retrograde ratio of initial distribution.
Though the octupole term from the solar perturbations can flip a very inclined system \citep{Naoz2011}, it also has no preference for the directionality of the system.
\citet{Schlichting2008} predicted a retrograde preference for binaries formed by the $L^2S$ capture method.
On the other hand, the gravitational collapse method of \citet{Nesvorny2010} favors orbits in the direction of disk clump rotation, which simulations (cited therein) generally show as prograde.
Three-body methods ($L^3$ and momentum exchange) have a much weaker inclination dependence. 
The current inclination distribution could therefore serve as a tracer for these various formation methods.
Because the sense of motion can be hard to determine for TNBs, only a few TNBs have published unambiguous inclinations.
However, for the 16 TNB orbits listed in Table \ref{tab:obs}, there is a 4:1 prograde preference.
\\\\
The $H'$ parameter introduced above is also preserved by KCTF (except for spin-orbit interactions), and Table \ref{tab:obs} also lists the $H'$ for those 16 known orbits.
The widest orbits ($a>5\%r_{Hill}$) are apparently stable, with their large $a$ and low inclinations keeping them stable even at high eccentricities.
These systems likely have thus been undergoing low-amplitude Kozai cycles for most of the history of the solar system.
Table \ref{tab:circ} lists the known TNB systems in near-circular orbits; these systems appear to represent the outer edge of the close circular population identified in our simulations.
As more sensitive high-resolution imaging systems come on line, an increasing number of these tight systems should be detected.
The remaining systems smaller than $5\%r_{Hill}$ have values of $|H'|$ at or below unity, implying that they all should have some amount of tidal evolution.
Most still have a larger $|H'|$ than the typical values of $|H'_{circ}|$ in Table \ref{tab:sims}, meaning they could have a range of physical properties and still not circularize.
\\\\
Two eccentric systems do have an $|H'|$ small enough to stand out, though: 2004 PB$_{108}$ and 2001 QC$_{298}$.
These two systems happen to be the only scattered-disk binaries by the DES classification \citep{Elliot2005,Osip2002} in Table \ref{tab:obs}, 
but we think it more likely that their apparent stability is due to their non spherical shapes.
Indeed, Table \ref{tab:sims} shows that if both of the objects in each system had a $J_2$ of at least 0.01, they could easily be stably eccentric for the lifetime of the solar system.
The other eccentric systems do not require $J_2$ to be stable, but it would do no harm to their stability.
\\\\
The three systems just wider, 2001 XR$_{254}$, Altjira, and 275809, all have values of $|H'|$ just larger than one.
Since they are all also in the classical Kuiper Belt, these systems have potentially had very little tidal evolution since their original formation as binaries.
Interestingly, they also have very similar systems masses and mutual orbit inclinations.
Future investigations could help determine if this is simply coincidence or an optimal point for stability.

\section{Conclusions}

KCTF can significantly transform the orbits of trans-neptunian binaries.
At least 90\% of random synthetic TNB systems survive 4.5 Ga of KCTF evolution.
A third to half of the surviving TNB systems decay to circular orbits at less than 1\% of their mutual Hill radius.
Some of these systems can have values of J/J' similar to impact-generated systems.
The remaining systems are stable being eccentric over the lifetime of the solar system.
All resulting systems preserve their initial prograde/retrograde preference.
\\\\
The inclusion of $J_2$ lowers the effectiveness of KCTF, but does not eliminate it, especially for rubble-pile objects.
In addition, $J_2$ creates an island of stability that allows otherwise unstable observed system to be in permanent eccentric orbits.
A slower initial rotation rate or 10:1 mass ratio also slightly lower the effectiveness of KCTF, but do not change the basic trends.
\\\\
The observed population of TNB orbits fits well to our simulations with $J_2$.
These simulations predict that, as high-resolution observational systems improve, a large number of TNBs will be detected with very tight, circular orbits.
Indeed, considering the fraction of known wider binaries, tight near equal-mass TNBs may be extremely common.

\section*{Acknowledgements}
This work was partially supported by grants HST-GO-11178 and HST-GO-11650 from the Space Telescope Science Institute, which is operated by the Association of Universities 
for Research in Astronomy, Incorporated, under NASA contract NAS5-26555.  
Partial support was also provided by NASA through Spitzer Space Telescope Cycle 5 grant RSA\#1353066 funded through a contract issued by the Jet Propulsion Laboratory, 
California Institute of Technology under a contract with NASA.  
Partial support was also provided by NSF Planetary Astronomy grant AST-1109872.
Special thanks to Frank Timmes and Travis Barman for contributing to the many CPU-hours it took to complete this project.
Thanks also to Darin Ragozzine and an anonymous reviewer for their helpful comments and suggestions.

\section{References}

\bibliography{poster}{}
\bibliographystyle{elsarticle-harv}

\begin{table*}[p!]
\begin{center}
\caption{Summary of KCTF simulation results: \% Circ. are the fraction of surviving orbits with $e<10^{-4}$,
\% Dest. are the fraction of initial orbits destroyed over the simulation, and $|H'_{tide}|$ and $|H'_{circ}|$
are as defined in the text.
}
\label{tab:sims}
\begin{tabular}{cccccc|cccc}
$Q$ & Spin & $J_2$ & $m_1/m_2$ & $k_L$ & $\rho$ & \% Circ. & \% Dest. & $|H'_{tide}|$ & $|H'_{circ}|$ \\
\hline
\hline
10  & fast &    0 &  1 & rubble & 0.5 &  56\% &  3\% & 1.28 & 0.620 \\
10  & fast &    0 &  1 & rubble & 1.0 &  48\% &  1\% & 1.19 & 0.550 \\
10  & fast &    0 &  1 &  solid & 0.5 &  38\% &  2\% & 1.05 & 0.435 \\
10  & fast &    0 &  1 &  solid & 1.0 &  39\% &  1\% & 1.03 & 0.426 \\
\hline
100 & fast &    0 &  1 & rubble & 0.5 &  48\% &  2\% & 1.07 & 0.534 \\
100 & fast &    0 &  1 & rubble & 1.0 &  42\% &  1\% & 0.98 & 0.488 \\
100 & fast &    0 &  1 &  solid & 0.5 &  29\% & 11\% & 0.92 & 0.380 \\
100 & fast &    0 &  1 &  solid & 1.0 &  34\% &  5\% & 0.94 & 0.324 \\
\hline
100 & fast &    0 & 10 & rubble & 0.5 &  43\% &  2\% & 1.05 & 0.437 \\
100 & fast &    0 & 10 & rubble & 1.0 &  40\% &  1\% & 1.04 & 0.359 \\
100 & fast &    0 & 10 &  solid & 0.5 &  23\% & 15\% & 0.84 & 0.326 \\
100 & fast &    0 & 10 &  solid & 1.0 &  25\% &  9\% & 0.87 & 0.264 \\
\hline
100 & fast &  0.1 &  1 & rubble & 0.5 &  38\% &  2\% & 1.12 & 0.005 \\
100 & fast &  0.1 &  1 & rubble & 1.0 &  36\% &  0\% & 1.11 & 0.005 \\
100 & fast &  0.1 &  1 &  solid & 0.5 &  24\% &  8\% & 0.90 & 0.012 \\
100 & fast &  0.1 &  1 &  solid & 1.0 &  28\% &  5\% & 0.84 & 0.001 \\
\hline
100 & fast & 0.01 &  1 & rubble & 0.5 &  36\% &  1\% & 1.09 & 0.028 \\
100 & fast & 0.01 &  1 & rubble & 1.0 &  34\% &  1\% & 1.04 & 0.054 \\
100 & fast & 0.01 &  1 &  solid & 0.5 &  25\% &  9\% & 0.87 & 0.014 \\
100 & fast & 0.01 &  1 &  solid & 1.0 &  26\% &  5\% & 0.90 & 0.014 \\
\hline
100 & slow &    0 &  1 & rubble & 0.5 &  44\% &  2\% & 1.13 & 0.524 \\
100 & slow &    0 &  1 & rubble & 1.0 &  43\% &  1\% & 1.00 & 0.514 \\
100 & slow &    0 &  1 &  solid & 0.5 &  27\% & 14\% & 0.88 & 0.364 \\
100 & slow &    0 &  1 &  solid & 1.0 &  33\% &  5\% & 0.94 & 0.362 \\
\end{tabular}
\end{center}
\end{table*}

\begin{table*}[p!]
\begin{center}
\caption{$H'$ for observed systems with fully-constrained orbits. $I$ is the angle between the heliocentric and mutual orbital planes, and $GM$ is the system mass.
Orbits are from \citet{Grundy2011}, \citet{Parker2011}, \citet{Sheppard2012} and references therein.}
\label{tab:obs}
\begin{tabular}{ll|ccccc}
Designation & Name & $a$ (\% $r_{Hill}$) & $e$ & $I$ (deg) & $GM$ (km$^3$/s$^2$) & $H'$ \\
\hline
\hline
79360           & Sila           &  0.35 & 0.02 & 123.1 & 0.72 & -0.32 \\
2001 QC$_{298}$ &                &  0.50 & 0.34 &  73.5 & 0.79 & +0.19 \\
66652           & Borasisi       &  0.91 & 0.47 &  49.4 & 0.23 & +0.55 \\
42355           & Typhon         &  1.15 & 0.53 &  50.5 & 0.06 & +0.58 \\
2004 PB$_{108}$ &                &  1.48 & 0.44 &  83.2 & 0.63 & +0.13 \\
2001 XR$_{254}$ &                &  1.70 & 0.56 &  21.1 & 0.27 & +1.01 \\
148780          & Altjira        &  1.83 & 0.34 &  25.4 & 0.27 & +1.15 \\
275809          &                &  1.88 & 0.42 & 161.0 & 0.27 & -1.18 \\
26308           &                &  2.42 & 0.47 &  75.4 & 0.46 & +0.35 \\
2003 QY$_{90}$  &                &  3.19 & 0.66 &  51.4 & 0.03 & +0.84 \\
58534           & Logos          &  3.25 & 0.55 &  74.2 & 0.03 & +0.41 \\
88611           & Teharonhiawako &  5.82 & 0.25 & 127.7 & 0.16 & -1.43 \\
\end{tabular}
\end{center}
\end{table*}

\begin{table*}[p!]
\begin{center}
\caption{Observed TNB systems with near-circular orbits. $GM$ is the system mass.
Orbits are from \cite{Brown2010}, \citet{Grundy2011}, \citet{Stansberry2011} and references therein.}
\label{tab:circ}
\begin{tabular}{ll|ccc}
Designation & Name & $a$ (\% $r_{Hill}$) & $e$ & $GM$ (km$^3$/s$^2$) \\
\hline
\hline
120347          & Salacia        &  0.23 & 0.01 & 30.28 \\
79360           & Sila           &  0.35 & 0.02 & 0.72 \\
90482           & Orcus          &  0.42 & 0.00 & 42.40 \\
134860          &                &  0.53 & 0.09 & 0.14 \\
123509          &                &  0.66 & 0.01 & 0.07 \\
65489           & Ceto           &  0.71 & 0.01 & 0.37 \\
\end{tabular}
\end{center}
\end{table*}

\end{document}